# On the universal scattering time of neutrons

*Guenter Nimtz[1] and Paul Bruney[2]*

[1] *II. Physikalisches Institut, Universität zu Köln, Germany*

[2] *Plane Concepts Inc., Silver Spring, Maryland, USA*

The tunneling and barrier interaction times of neutrons have been previously measured. Here we show that the neutron interaction time with barriers corresponds to the universal tunneling time of wave mechanics, which was formerly observed with elastic, electromagnetic, and electron waves. The universal tunneling time seems to also hold for neutrons. Such an adequate general wave mechanical behavior was conjectured by Brillouin. Remarkably, wave mechanical effects and even virtual particles hold from the microcosmos up to the macrocosmos.

Neutron tunneling and neutron interaction times have been recently studied (1-3). As expected by Brillouin in his textbook *Wave propagation in periodic structures*, the waves of all fields should act in a similar way. Here we evidence this speculation in the case of tunneling.

Photonic tunneling time was studied in various barrier systems and at different electromagnetic wave frequencies. The comparison of the available experimental data revealed a universal tunneling time (4, 5). The tunneling time was found to be approximately equal the reciprocal photon or phonon frequency

$$\tau \approx 1/\nu = h/E, \qquad\qquad 1$$

where τ is the measured group tunneling time, ν the radiation frequency, h the Planck constant and E the wave packet's energy. As shown in Tab. 1, it was observed that besides photons, phonons and electrons also exhibit such an approximate universal tunneling time.

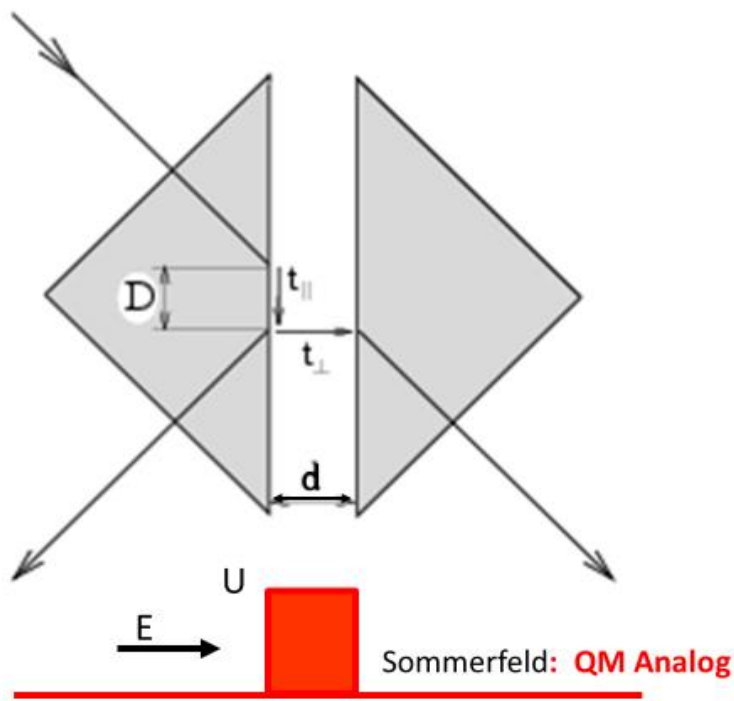

Fig. 1 Double prism, the optical QM tunneling analog according Sommerfeld (8). E and U are the particles energy and the potential barriers height. D are the Goos-Hänchen shift along the surface and d the width of the gap between the two prisms respectively (9). The gap traversal time $t_⊥$ is observed to be zero and $t_∥$ represents the effective tunneling time τ. In such a symmetric set-up both the reflected and the transmitted beams are detected at the same time.

Esposito (6) and somewhat later Olkhovsky et al. (7) delivered a theoretical description of this universal tunneling time for rectangular barriers. Their theoretical data fits the experimental photonic values in the first order approximation (6, 7)

| Tunneling barriers | τ | T=1/ν |
|---|---|---|
| Frustrated total reflection | 117 ps | 120 ps |
| Double prisms | 87 ps | 100 ps |
| Photonic lattice | 2.13 fs | 2.34 fs |
| Photonic lattice | 2.7 fs | 2.7 fs |
| Undersized waveguide | 130 ps | 115 ps |
| Electron field-emission tunneling | 7 fs | 6 fs |
| Electron ionization tunneling | ≤ 6 as | ? as |
| Acoustic (phonon) tunneling | 0.8 µs | 1 µs |
| Acoustic (phonon) tunneling | 0.9 ms | 1 ms |
| Neutron (resonant level) | 0.217 µs | 0.236 µs* |
| Neutron (non-resonant) | > 19 ns | 33 ns ** |

Table 1: τ are examples of measured time and T is the reciprocal frequency (5). Data * is a calculated phase time value (7). Data ** from Eq.1 with 127 neV.

Quite recently, Mativane et al. measured the reflectivity of tunneling neutrons, i.e. of Schrödinger waves (2). The experiments were carried out in grazing angle geometry. That is, in the angle range of frustrated total internal reflection as this reflection is called in optics. The double prism is the quantum mechanical analog of the tunneling process. Total reflection becomes frustrated if the second prism approaches the first one and light is partially transmitted to the second prism (9). In this neutron experiment a sandwich nanostructure of $^{58}$Ni$^{62}$Ni$^{58}$Ni layers represents a Fabry-Perot resonator. The two Ni isotopes have a different neutron scattering potential. Seven tunneling resonances of this resonator have been detected in the reflectivity spectrum (2).

The so-called thickness or Kiessig fringes are seen at larger angles and shorter wavelengths. The period of these fringes gives an approximate value of the resonator thickness.

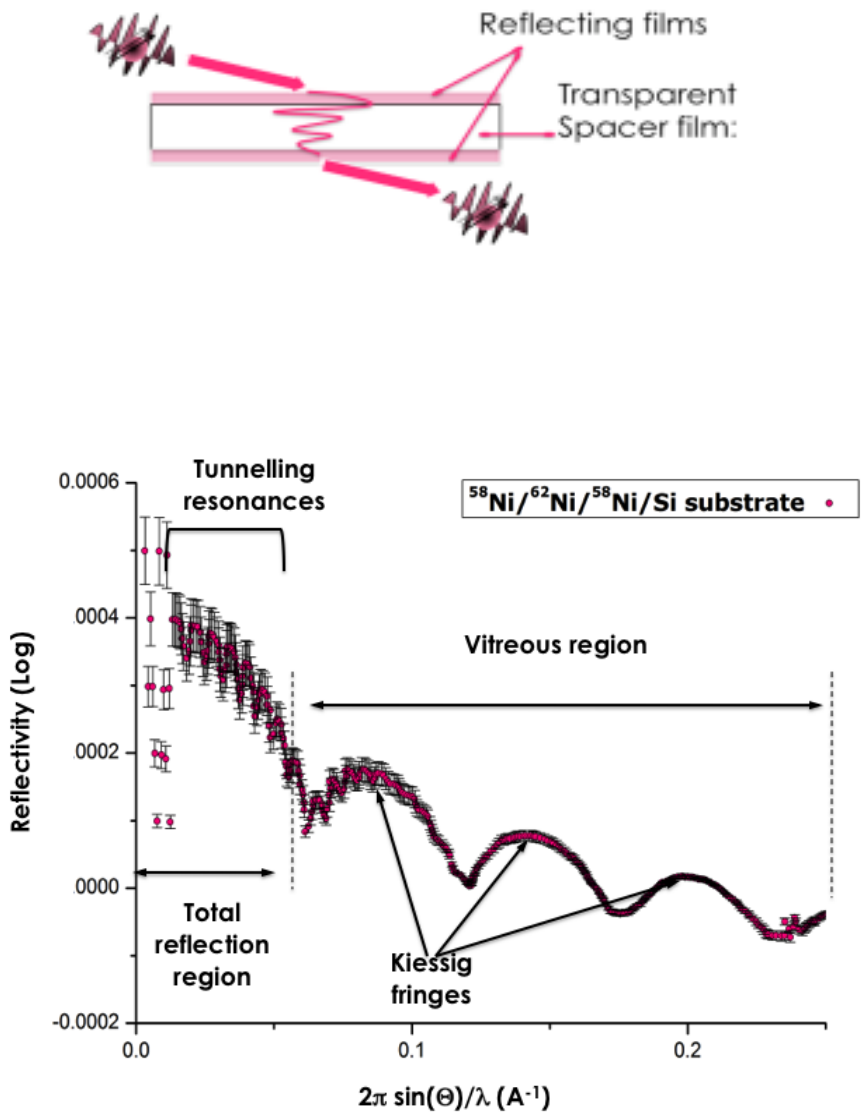

Fig.2  On top an illustration of the grazing angle interaction at thin films. Below, reflectivity by resonant tunneling on a $^{58}Ni/^{62}Ni/^{58}Ni$ Fabry-Perot sandwich structure of two Ni isotopes, where θ is the angle between surface and incident neutron beam and λ the interacting neutron wave length (2).

In the interaction experiments of Frank et al. (1), the Bragg reflection time was studied with neutrons of ≈ 2 nm wavelength. The barrier structures were built from 30 pairs of a periodic thin film structure of 13.0 nm thick Ni-V alloy and 7.0 nm thick Ti layers based on a Si substrate. Grazing incident neutron beams were applied. An average interaction time over the measured grazing angle range of the order of 0.2 µs was measured. Assuming the double Bragg reflection is due to the Ni-V alloy Ti layers, a single peak time was 0.156 µs. This time corresponds with Eq. 1 to a reflected energy of 26.6 neV, assuming the Bragg interaction as due to a periodic lattice.

The energy of a resonant level of a sandwich structure by Ni --- Ti/Zr alloy --- Ni layers was estimated to be ≈ 127 neV. The two Ni layers present a double barrier separated by the Ti-Zr alloy. From Eq. 1, a tunneling time of 33 ns is obtained for a non-resonant tunneling. However, a resonant level tunneling time is always much longer as it represents a cavity (10). Olkhovsky et al. (7) have calculated the value given in the table for the interaction time with the resonant level. A different structure was studied by Maaza et al. (3) they observed 5 levels and estimated lifetimes of the resonant neutron waves between 0.1 and 1 µs.

The above value of 33 ns approximately equals the measured value near the resonance and was assumed to present the travel time through the Si substrate.

Summing up, elastic, electromagnetic, and Schrödinger waves reveal an approximative universal potential barrier scattering time. The effect is observed in the tunneling process and in the Bragg diffraction time. It is described by the Wigner phase time and the Helmholtz and Schrödinger equations, respectively. Additionally, the universal tunneling time agrees with the theoretical study of Hartman (11). At that time he was motivated by the first tunneling experiments in solid state physics by Esaki's tunneling diode and by Giaever's tunneling between two superconductors separated by a thin metal oxide layer. The interaction and tunneling time arises at the barrier front. The non-local propagation inside a barrier is described by evanescent modes and virtual wave packets (12, 13).

Acknowledgement: We are very grateful to Prof. M. Maaza for his support with figure 2.